\documentclass[a4paper,fleqn,usenatbib]{mnras}

\usepackage[T1]{fontenc}
\usepackage{ae,aecompl}

\usepackage{graphicx}	% Including figure files
\usepackage{amsmath}	% Advanced maths commands
\usepackage{amssymb}	% Extra maths symbols
\usepackage{txfonts}

\title[Polarization of Scattered Radiation]{Inclination-Induced Polarization of Scattered 
  Millimeter Radiation from Protoplanetary Disks: The Case of HL Tau}

\author[H. Yang et al.]{
Haifeng Yang,$^{1}$\thanks{E-mail: hy4px@virginia.edu}
Zhi-Yun Li,$^{1}$
Leslie Looney$^{2}$
and Ian Stephens$^{3}$
\\
$^{1}$Astronomy Department, University of Virginia, Charlottesville, VA 22904, USA\\
$^{2}$Department of Astronomy, University of Illinois, Urbana, Illinois 61801, USA\\
$^{3}$Institute for Astrophysical Research, Boston University, Boston, MA 02215, USA
}

\date{Accepted XXX. Received YYY; in original form ZZZ}

\pubyear{2015}

\begin{document}
\label{firstpage}
\pagerange{\pageref{firstpage}--\pageref{lastpage}}
\maketitle

\begin{abstract}

Spatially resolved polarized millimeter/submillimeter emission 
has been observed in the disk of HL Tau and two other young 
stellar objects. It is usually interpreted as coming
from magnetically aligned grains, but can also 
be produced by dust scattering, as demonstrated explicitly 
by Kataoka et al. for face-on disks. We extend their work 
by including the polarization induced by disk inclination 
with respect to the line of sight. Using a physically 
motivated, semi-analytic model, we show that the polarization 
fraction of the scattered light increases with the inclination 
angle $i$, reaching $1/3$ for edge-on disks. The inclination-induced 
polarization can easily dominate that intrinsic to the disk in 
the face-on view. It provides a natural explanation for the two main
features of the polarization pattern observed 
in the tilted disk of HL Tau ($i \sim 45^\circ$): the polarized 
intensity concentrating in a region elongated more or less along 
the major axis, and polarization in this region roughly parallel 
to the minor axis. This broad agreement provides support to dust
scattering as a viable mechanism for producing, at least in part, 
polarized millimeter radiation. In order to produce polarization 
at the observed level ($\sim 1\%$), the scattering grains 
must have grown to a maximum size of tens of microns. However, such 
grains may be too small to produce the opacity spectral index of 
$\beta \lesssim 1$ observed in HL Tau and other sources; another 
population of larger, millimeter/centimeter-sized, grains may be
needed to explain the bulk of the unpolarized continuum emission. 

\end{abstract}


%%%%%%%%%%%%%%%%%%%%%%%%%%%%%%%%%%%%%%%%%%%%%%%%%%

%%%%%%%%%%%%%%%%% BODY OF PAPER %%%%%%%%%%%%%%%%%%

\section{Introduction}
\label{sec:intro}

\begin{figure}
  \includegraphics[width=0.49\textwidth]{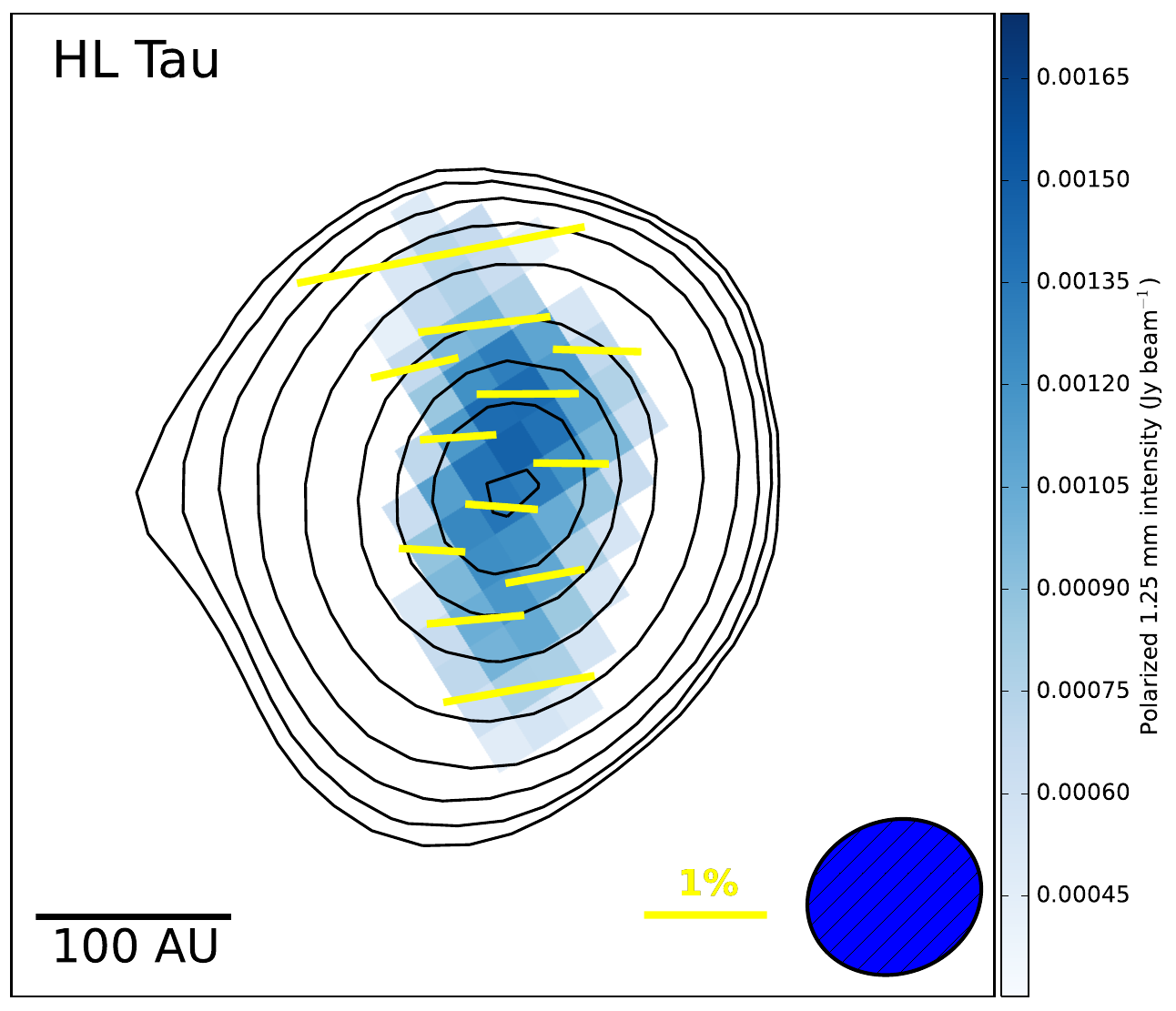}
  \caption{Observed 1.3 mm polarization pattern of the HL Tau
    disk. Plotted are total intensity (contours), polarized intensity
    (color map, $> 2\sigma$), and polarization vectors ($>3\sigma$), with length
    proportional to the polarization fraction. Adapted from Stephens et
    al. (2014). The major axis determined from the ALMA data (ALMA
    Partnership: Brogan et al. 2015) is tilted
    by $10^\circ$ clockwise with respect to the vertical direction.   
  } \label{fig:obs}
\end{figure}

Polarized millimeter/sub-millimeter emission has been observed in the 
disks around 3 young stellar objects: IRAS 16293-2422B \citep{rao2014}, 
HL Tau \citep{stephens2014} and L1527 \citep{segura-cox2015}. 
It is usually interpreted as coming from magnetically aligned 
dust grains (e.g., \citealt{cl2007}). This interpretation appears 
consistent with the data on IRAS 16293-2422B and L1527,  where the 
observed polarization patterns are broadly similar to those expected 
from grains aligned by a predominantly toroidal magnetic field 
(\citealt{rao2014, segura-cox2015}; see \S~\ref{sec:discussion} for 
more discussion). In contrast, the polarization 
vectors in the disk of HL Tau (which is inclined with respect to 
the line of sight by $\sim 45^\circ$, \citealt{alma2015, kwon2011})
are all roughly parallel to the minor axis 
(see Fig.\ref{fig:obs}), which is not compatible with the pattern 
expected of grains aligned by a toroidal disk field, although 
more complicated field configurations cannot be ruled out \citep{stephens2014}. 
Another drawback of the toroidal field-aligned grain 
model is that it predicts a lower polarization fraction along 
the major axis than along the minor axis \citep[see][Fig.~10]{cl2007}, 
which is the opposite of what is observed 
(Fig.~\ref{fig:obs}). A much better fit is provided by a 
uni-directional magnetic field approximately (within $\sim 10^\circ$) 
along the disk major axis. However, such a field configuration would 
be difficult to maintain against the disk differential rotation. 
Furthermore, there is growing evidence for grain growth in
protoplanetary disks, up to millimeter or even centimeter 
sizes (e.g., \citealt{perez2012, alecian2013, testi2014}). 
It is unclear whether such large grains would be aligned with 
respect to the magnetic field through the currently favored 
mechanism of radiative torque because their magnetic moments 
may not be large enough to provide the fast precession needed 
\citep{lazarian2007} and their slow internal relaxation makes the 
alignment less efficient \citep{HL2009}.  
These difficulties provide a motivation to investigate other potential 
mechanisms for producing, at least in part, the polarized mm emission 
from the HL Tau disk.

One possibility, first investigated in detail by \cite{kataoka2015a}, 
is that the polarized millimeter disk emission comes from
scattering of large dust grains. Kataoka et al. stressed the 
need for anisotropic radiation field for this mechanism to work
efficiently. As examples of this requirement, they considered
radiation from structured disks with either an axisymmetric ring 
or a non-axisymmetric lobe, both viewed face-on. Here, we extend 
\citeauthor{kataoka2015a}'s work to show that significant polarization is 
produced even in a smooth disk, as long as the disk axis is 
tilted away from the line of sight by a large enough angle. The 
physical reason can be understood most easily in the limits of
optically and geometrically thin dust emission and Rayleigh
scattering, where the degree of polarization is peaked near a
scattering angle of $90^\circ$. In these limits, the radiation field
to be scattered by dust grains becomes essentially two 
dimensional, concentrated in the disk 
plane. In such a case, only the radiation propagating
along the disk major axis would be scattered by $90^\circ$ to reach
the observer and be maximally polarized, with a polarization direction
along the minor axis in the plane of the sky. This simple geometric
effect provides a natural explanation for the two main features 
of the polarization pattern observed in the HL Tau disk (see 
Fig.~\ref{fig:obs}): (1) polarized emission concentrating in an 
elongated region more or less along the major axis, and (2) 
polarization vectors in this region roughly parallel to the 
minor axis. These are the features that are difficult to explain in
the current models of polarization from magnetically aligned
grains\footnote{Grains aligned by a toroidal magnetic field can 
explain the orientations of the polarization vectors near the 
center, but not at larger distances (see the left panel of 
Fig.~2 of \citealt{stephens2014}).}. They provide a strong motivation 
to explore in more detail the polarization pattern produced by 
dust scattering. It is a first step toward disentangling the 
contributions of dust scattering and grain alignment to the 
disk polarization, which is needed in order to take full advantage of
the high resolution ALMA and JVLA polarization observations that will
become available soon for probing the grain growth and/or magnetic
field in protoplanetary disks. 

In the remainder of the paper, we will put the above qualitative 
picture on a more quantitative ground by computing the scattered 
radiation from an inclined disk (such as the HL Tau disk) 
semi-analytically using the so-called ``thin-disk'' approximation 
that brings out the essential physics transparently 
(\S~\ref{sec:polarized}); a more complete treatment of the 
radiative transfer problem will be postponed to a future 
investigation. The effects of disk inclination on the polarization
pattern will be emphasized. The scattered radiation is compared with the direct dust emission in \S~\ref{sec:application}, where the 
fractional polarization of the total intensity is computed. In
\S~\ref{sec:implications}, 
we comment on the implications of the polarization fraction observed 
in HL Tau on the grain size distribution. We discuss the possibility 
of dust scattering-induced polarization for other sources and 
future model refinements in \S~\ref{sec:discussion} and 
conclude in \S~\ref{sec:conclusion}. 

\section{Polarized Radiation from Dust Scattering}
\label{sec:polarized}

In this section, we develop a semi-analytic model for polarization
from dust scattering under the simplifying assumptions that the dust
emission is optically thin and that the dust disk is geometrically
thin, with a local scale height much less than the radius. The 
latter would be a particularly good approximation if the relatively 
large grains responsible for the emission and scattering of 
millimeter radiation are settled toward the disk mid-plane. To
isolate the polarization due to dust scattering from that caused by
grain alignment, we further assume that the grains are not aligned, so
that the radiation directly emitted by dust is unpolarized. Under 
these assumptions, the incoming radiation field seen by the
scattering dust grains at a given point inside the disk can be
decomposed into two components: a local, roughly isotropic, component
emitted by the (non-aligned) grains in a region of size comparable to 
the (dust) scale height surrounding the point, and an anisotropic 
component coming from the grains in the rest of the disk beyond the 
local region.  The former does not lead to significant polarization 
after scattering because scattering of isotropic radiation by
non-aligned grains does not have any preferred direction. It is the 
latter that is mainly responsible for the polarized radiation 
after scattering. The computation of this component will be 
drastically simplified by the thin-disk approximation, as we show 
below. 

\subsection{Formulation of the Problem}
\label{subsec:formulation}

To facilitate the computation of the scattered light from dust grains
at a location ${\bf r}$ inside a disk with axis 
inclined by an angle $i$ with respect to the line of sight, we 
define two coordinate systems, both centered on the scatterer's 
location ${\bf r}$. The first is fixed in the observer's frame, with 
horizontal $x'$- and vertical $y'$- axes in the plane of the sky, and 
$z'$-axis pointing toward the observer. The second is fixed on the 
disk, with the $y$-axis coinciding with the $y'$-axis, and the 
$z$- and $x$-axis rotating around the $y'$-axis counter-clockwise 
(viewed along the minus $y'$-axis) from the $z'$- and $x'$-axis, 
respectively, by the inclination angle $i$. We align the major axis of the 
disk in the plane of the sky along the $y$- (and $y'$-) axis (see 
Fig.~\ref{fig:illustration} below for an illustration), and the disk normal 
direction along the $z$-axis. 

Since the scattered radiation is linearly polarized, we will compute the
three Stokes parameters $I_s$, $Q$ and $U$ defined in the observer's 
frame $(x',y',z')$ separately, starting with the intensity $I_s$ (we 
drop the subscript $s$ for $Q$ and $U$ since, in our model, they are 
assumed to be produced only by the scattered radiation, without any
contribution from the direct radiation, unlike the total intensity $I$). 
In the optically thin limit, the intensity of the scattered radiation 
along the line of sight $I_s$ is given by an integration of the source
function of the scattered radiation, $S_{s,z'}$, along the $z'$-axis. The
source function $S_{s,z'}$ at any location ${\bf r}$ inside the disk is 
given by  
\begin{equation}\label{eq:source_function}
S_{s,z'} ({\bf r}) = \frac{1}{\sigma_s}\int \frac{d\sigma}{d\Omega}
I({\bf r},\theta, \phi)\, d\Omega
\end{equation}
where $I({\bf r}, \theta,\phi)$ is the intensity of the unpolarized
incoming radiation seen by the scattering dust
grains at the location ${\bf r}$ along the direction of polar angle $\theta$ 
(measured from the $z$-axis or the disk normal direction) and azimuthal 
angle $\phi$ (measured counter-clockwise from the $x$-axis on the
disk), $\sigma_s$ is the 
solid angle $\Omega$-integrated (total) scattering cross section,  
and $d\sigma/d\Omega$ is the differential cross section for scattering 
the radiation $I({\bf r},\theta,\phi)$ into the line of sight (i.e., 
along the $z'$-axis). 

The intensity $I({\bf r},\theta,\phi)$ is given by an integration of 
the source function for thermal dust emission, the Planck 
function $B_\nu(T)$, over the optical depth
$d\tau_{abs}=\kappa_{abs}\rho\, dl$ (where $\kappa_{abs}$ is the thermal 
dust absorption opacity and $\rho$ the mass density 
at a source location ${\bf r_l}$) along a line in the direction 
$(\theta,\phi)$ up to the scatterer at ${\bf r}$ (the quantity $l$ 
is the distance between ${\bf r_l}$ and ${\bf r}$):
 \begin{equation}\label{eq:intensity}
I({\bf r},\theta, \phi) = \int_0^\infty \kappa_{abs}({\bf r_l}) \frac{2\nu^2 kT
  ({\bf r_l})}{c^2} \rho({\bf r_l})\, dl,
\end{equation}
where we have assumed that the photon energy $h\nu$ is substantially 
less than $kT$ so that the Rayleigh-Jeans law $B_\nu(T)=2\nu^2 kT/c^2$
is applicable; we have checked that, for the HL Tau model to be discussed 
below in \S~\ref{disk_model}, the results will not change
significantly if the Planck function is used instead. 

Substituting the above expression for $I({\bf r},\theta, \phi)$ into
Eq.\eqref{eq:source_function} and reorder the integrals, we can
rewrite the source function for the scattered radiation into:
\begin{equation}\label{eq:source2}
  \begin{split}
S_{s,z'} =& \frac{2\nu^2k}{c^2\sigma_s}\int_0^{2\pi}d\phi \int_0^\infty
dl \int_0^\pi d\theta \frac{d\sigma}{d\Omega} \kappa_{abs}({\bf r_l})\rho({\bf r_l}) T({\bf r_l}) \sin\theta\\
\equiv& S_0 + S_\infty,
  \end{split}
\end{equation}
where 
\begin{equation}\label{eq:S_0_def}
S_0\equiv \frac{2\nu^2k}{c^2\sigma_s}\int_0^{2\pi}d\phi \int_0^H
dl \int_0^\pi d\theta \frac{d\sigma}{d\Omega} \kappa_{abs}({\bf
  r_l})\rho({\bf r_l}) T({\bf r_l}) \sin\theta,  
\end{equation}
and 
\begin{equation}\label{eq:S_infty_def}
S_\infty\equiv \frac{2\nu^2k}{c^2\sigma_s}\int_0^{2\pi}d\phi \int_H^\infty
dl \int_0^\pi d\theta \frac{d\sigma}{d\Omega} \kappa_{abs}({\bf r_l})
\rho({\bf r_l}) T({\bf r_l}) \sin\theta. 
\end{equation} 
The two quantities, $S_0$ and $S_\infty$, denote the contributions to 
the unpolarized incoming radiation to be scattered at the location 
${\bf r}$ from two conceptually distinct regions respectively. 
For a geometrically thin dust disk, the unpolarized radiation 
coming from a region within a distance 
on the order of the local (dust) scale height $H$ is expected to 
be more or less isotropic. This near-field contribution, denoted 
by $S_0$, produces little polarized radiation after scattering; it 
will be discussed in the next section, together with the unpolarized 
direct dust emission. 
In contrast, the thermal dust emission coming from well outside this 
local region (i.e., $l >> H$) is mostly confined close to the disk 
plane. This far-field contribution, denoted by $S_\infty$, is highly 
anisotropic, with the bulk of the radiation beamed into a narrow 
range of polar angle near $\theta=\pi/2$ along any azimuthal 
direction $\phi$. Specifically, the vertical 
column of disk material passing through the source location 
${\bf r_l}$ at a distance $l$ from the scatterer contributes 
radiation only within a range of polar angle  
$\delta\theta \sim 2 H({\bf r_l})/l << 1$, where $H({\bf
  r_l})$ is the (dust) scale height at ${\bf r_l}$. 
Replacing the integral over angle $\theta$ in the expression for 
$S_\infty$ (eq.~\eqref{eq:S_infty_def}) by this rough estimate 
and making use of $\theta\approx \pi/2$, we have approximately 
 \begin{equation}\label{eq:S_infty}
 S_\infty\approx \frac{2\nu^2 k\kappa_{abs}}{c^2\sigma_s}\int_0^{2\pi}d\phi
 \frac{d\sigma}{d\Omega} \Lambda({\bf r},\phi),
\end{equation}
 where we have assumed a spatially constant absorption opacity
 $\kappa_{abs}$ for simplicity. The auxiliary quantity 
$\Lambda({\bf r},\phi)$ is a 
line integral along the direction of constant $\phi$ in the disk 
plane defined as
 \begin{equation}\label{eq:densityIntegral}
 \Lambda({\bf r},\phi) \equiv \int_H^\infty dl \frac{\Sigma({\bf
     r_l})T({\bf r_l})}{l},
\end{equation}
where $\Sigma({\bf r_l})=2\rho({\bf r_l}) H({\bf r_l})$
is the column density at the source location ${\bf r_l}$. 

One can determine the distance between the source location ${\bf
  r_l}$ and the center of the disk through 
\begin{equation}\label{eq:distance_to_center}
r_l = \sqrt{r^2+l^2 - 2 r l \cos(\phi-\phi_{\bf r})},
\end{equation}
where $r$ and $\phi_{\bf r}$ are the radius and azimuthal angle of the
scatterer in an $(x,y,z)$ coordinate system that centers on the
star (rather than the scatterer at ${\bf r}$). For the axisymmetric 
disk that we will consider below, the radius $r_l$ uniquely determines 
the column density $\Sigma$ and temperature $T$ that appear in 
eq.\eqref{eq:densityIntegral}. 

Once the integral $\Lambda({\bf r},\phi)$ is computed, the only 
term that is left to determine in eq.\eqref{eq:S_infty} is
the differential cross section for scattering $d\sigma/d\Omega$. 
For illustrative purposes, we will consider the dust scattering 
under the Rayleigh 
approximation, which is valid when the grain sizes are smaller 
than the wavelength divided by $2\pi$ (see, e.g., Fig.~10 of 
\citealt{canovas2013} and Fig.~\ref{fig:preversal} below); we will 
check in \S~\ref{subsec:grain} that 
this condition is satisfied for the HL Tau model to be discussed 
in the next section. In this limit, the differential cross section 
is given by
\begin{equation}\label{eq:cross_section}
\frac{d\sigma}{d\Omega} = 
\frac{3\sigma_s}{16\pi}\left(1+\cos^2\theta_s\right), 
\end{equation}
where $\theta_s$ is the scattering angle between the incoming
radiation along 
the direction $(\theta,\phi)$ and the outgoing radiation along the 
line of sight (i.e., the $z'$-axis), given by
\begin{equation}\label{eq:scattering_angle}
\cos\theta_s=\sin i \cos \phi.
\end{equation}
After scattering, a fraction of the initially unpolarized incoming 
radiation becomes polarized. The polarization fraction is 
\begin{equation}\label{eq:pol_frac}
p = \frac{1-\cos^2\theta_s}{1+\cos^2\theta_s},
\end{equation}
which peaks at $\theta_s=\pi/2$. 

The polarization direction is perpendicular to the scattering plane 
formed by the incoming direction $(\theta,\phi)$ and the $z'$-axis. 
It lies in the $x'$-$y'$ plane of the sky, at an angle $\phi'+\pi/2$ 
measured counter-clockwise from the $x'$-axis, with the angle $\phi'$ 
given by 
\begin{equation}\label{eq:stokes_rotation_angle}
\cos \phi'=\frac{\cos i \cos \phi}{\sqrt{\sin^2\phi + \cos^2 i \cos^2 \phi}}.
\end{equation} 

This linearly polarized radiation is the source of the observed 
polarized radiation in the plane of the sky, through the source 
functions for the Stokes parameter $Q$ and $U$: 
 \begin{equation}\label{eq:S_Qinfty}
 S_{Q,\infty}\approx - \frac{2\nu^2 k\kappa_{abs}}{c^2\sigma_s}\int_0^{2\pi}d\phi
 \frac{d\sigma(\theta_s)}{d\Omega} \Lambda({\bf r},\phi)\; p(\theta_s)\; \cos(2\phi'),   
\end{equation}
and 
 \begin{equation}\label{eq:S_Uinfty}
 S_{U,\infty}\approx - \frac{2\nu^2 k\kappa_{abs}}{c^2\sigma_s}\int_0^{2\pi}d\phi
 \frac{d\sigma (\theta_s)}{d\Omega} \Lambda({\bf r},\phi)\;  
p(\theta_s)\; \sin(2\phi').  
\end{equation}
For the optically thin radiation that we are considering, $Q$ and $U$
along the line of sight passing through any location ${\bf r}$ in 
the disk is simply given by their respective source functions,
multiplied by the optical depth for scattering $\Delta\tau_s \approx 
\kappa_{sca} \Sigma({\bf r})/\cos i$ (where $\kappa_{sca}$ is the
scattering opacity) through the disk: $Q=S_{Q,\infty} \Delta\tau_s$ and 
$U=S_{U,\infty} \Delta\tau_s$. 

\subsection{Inclination-Induced Polarization}
\label{sec:isotropic}

As a simple illustration of polarization from scattered light 
induced by disk inclination, we consider the limiting 
case where the incoming radiation field seen by the scatterer at 
the location ${\bf r}$ is confined to an infinitely thin 
disk plane (so that the near-field contribution to the scattering 
source function, $S_0$, can be ignored compared to the far-field 
contribution $S_\infty$) and is 
isotropic in the azimuthal ($\phi$) direction (but highly anisotropic 
in polar angle $\theta$). In this case, the integral 
$\Lambda({\bf r},\phi)$ defined in eq.~\eqref{eq:densityIntegral} is
independent of $\phi$ and can be moved 
outside the integral over $\phi$ in the source functions for the 
total scattered intensity $I_s$ (eq.~\eqref{eq:S_infty}), Q
(eq.~\eqref{eq:S_Qinfty}) and U (eq.~\eqref{eq:S_Uinfty}), so that 
\begin{equation}\label{eq:S_limiting}
  \begin{split}
S_\infty &= \frac{2\nu^2 k\kappa_{abs} \Lambda}{c^2} \int_0^{2\pi} d\phi
\frac{3}{16\pi}(1+\sin^2i\cos^2\phi) \\
&= C\int_0^{2\pi} d\phi (1+\sin^2i\cos^2\phi) = \pi C(2+\sin^2 i),
  \end{split}
\end{equation}
where $C=3\nu^2 k \kappa_{abs} \Lambda/(8\pi c^2)$ is a constant independent 
of the angles $i$ and $\phi$, and 
\begin{equation}\label{eq:S_Qlimiting}
  \begin{split}
S_{Q,\infty} &= - C \int_0^{2\pi} d\phi (1-\sin^2i\cos^2\phi)
\frac{\cos^2\phi\cos^2i-\sin^2\phi}{\cos^2\phi\cos^2i+\sin^2\phi} \\
&= \pi C \sin^2 i,
  \end{split}
\end{equation}
\begin{equation}\label{eq:S_Ulimiting}
S_{U,\infty} = -C \int_0^{2\pi} d\phi (1-\sin^2i\cos^2\phi) 
\frac{\cos i\sin(2\phi)}{\cos^2\phi\cos^2i+\sin^2\phi} = 0.
\end{equation}

The fact that $S_{U,\infty}$ is zero and $S_{Q, \infty}$ is positive
(for $i\neq 0$) means that the inclination-induced polarization 
is always along the $x'$-axis (or the minor axis of the disk) 
in the plane of the sky for Rayleigh scattering 
(see Fig.~\ref{fig:illustration} for the $x'$-$y'$ coordinates, 
and eq.~\eqref{eq:polarization_angle} below for the
relation between the Stokes parameters $Q$ and $U$, and polarization 
angle $\alpha$). This is expected physically because,   
in a tilted disk, the light coming from different directions in
the disk plane will be scattered by different angles toward 
the observer. In particular, the light coming from a direction 
along the major axis will always be scattered by $\pi/2$. In the 
Rayleigh limit, this part of the light will be fully polarized 
along the minor axis of the disk. In contrast, the light coming 
from a direction along the minor axis will be scattered by 
either $\pi/2-i$ or $\pi/2+i$. This part of the light will be 
partially polarized along the major axis, with a polarization 
fraction of $\cos^2i/(1+\sin^2i)$. The difference in the fraction  
of polarization leads to more scattered light polarized along 
the minor axis than along the major axis, despite the fact 
that the scattering cross section is less for the former than 
the latter (see eq.~\eqref{eq:cross_section}). This generic
tendency for the inclination-induced polarization to 
align with the minor axis provides a natural explanation 
for the polarization directions observed in HL Tau 
(see Fig.~\ref{fig:obs}). 

It is easy to determine the fraction of the total scattered light 
that is polarized along the minor axis:
\begin{equation}\label{eq:fraction_limiting}
    p_s(i) \equiv \frac{\sqrt{Q^2+U^2}}{I_s}=
  \frac{\sqrt{S_{Q,\infty}^2+S_{U,\infty}^2}}{S_\infty} 
 = \frac{\sin^2i}{2+\sin^2i}.
\end{equation}
This same expression can be obtained if one considers only the radiation
coming from directions along the major and minor axes. Note that the 
maximum degree of polarization reaches $1/3$ when the disk is viewed
edge-on ($i=90^\circ$). For the inclination angle $i=0^\circ$, $30^\circ$, 
$45^\circ$ and $60^\circ$ to be considered in the next subsection, 
the fractional polarization is $p_s=0$, $1/9$, $1/5$, and $3/11$, 
respectively. The analytically obtained polarization fraction in 
this simple limiting case will be used to interpret the results 
obtained numerically in more general cases.     
 
\subsection{Intrinsic Polarization from Azimuthally Anisotropic 
Radiation: an Example} 
\label{disk_model}

As emphasized by \cite{kataoka2015a}, the radiation field in the
disk plane is not isotropic in general, and the anisotropy in the 
azimuthal ($\phi$) direction leads to polarized 
scattered light even in the face-on case. In a tilted disk, the 
observed polarization pattern is expected to be shaped by the 
interplay between those produced by anisotropy in $\phi$-direction 
and disk inclination (which relies on strong anisotropy 
in $\theta$ direction). To illustrate this interplay, we adopt \cite{kwon2011}
model of the HL Tau disk, where the distributions 
of temperature and column density as a function of the cylindrical 
radius $R$ are parametrized as 
\begin{equation}\label{eq:temperature}
T=T_0\left(\frac{R}{R_c}\right)^{-q},
\end{equation}
\begin{equation}\label{eq:column_density}
\Sigma = \Sigma_0 \left(\frac{R}{R_c}\right)^{3/2-p-q/2} 
\exp\left[-\left(\frac{R}{R_c}\right)^{7/2-p-q/2}\right],
\end{equation} 
where $R_c$ is a characteristic disk radius beyond which the 
column density drops off exponentially, and $p$ is an exponent that,
together with the exponent $q$, controls the column density 
distribution; it is not to be confused with the polarization
fraction. The temperature 
profile yields a thermal scale height for the gas 
\begin{equation}\label{eq:scale-height}
H=H_0 \left(\frac{R}{R_c}\right)^{3/2-q/2}. 
\end{equation}
Although higher resolution ALMA observations have revealed substructures
(rings) on the disk \citep{alma2015}, the 
above is still the best model at the CARMA resolution that was used 
to detect the polarization in HL Tau.   
Adopting $q=0.43$, \citeauthor{kwon2011} found a set of parameters that best fit
their CARMA observations at 1.3 and 2.7~mm: 
$R_c\approx 79$~AU, $p\approx 1$, and a
scale height for gas at $R_c$ of $H_0\approx 16.8$~AU (see also \citealt{kwon2015}). 
Since the relatively
large grains responsible for the scattered radiation may settle
toward the midplane, the scaling factor $H_0$ in
eq.\eqref{eq:scale-height} may need to be reduced by some (potentially
large) factor \citep{kwon2011}. We have experimented with reduction 
factors of $1$, $10$, $50$ and $100$, and found very similar results. 
In what follows, we will focus on the case where the scale height is 
the same for the dust and gas.  

With the disk structure specified, we can now compute the Stokes
parameters $Q$, $U$ and $I_{s,\infty}$ from their source functions
given by eqs.~\eqref{eq:S_Qinfty}, \eqref{eq:S_Uinfty} and \eqref{eq:S_infty}. In this section, we consider 
the contribution $I_{s,\infty}$ to the total observed intensity from 
the far-field scattering source function $S_\infty$ only, in order to 
facilitate comparison with the analytic results obtained in the
preceding subsection; 
the contributions from $S_0$ and direct dust emission would lower 
the polarization fraction, and will be considered in the next 
section. From these three Stokes parameters, we can determine the total 
intensity of the polarized radiation 
\begin{equation}\label{eq:polarized_intensity}
I_p=\sqrt{Q^2+U^2},
\end{equation}
the polarization angle $\alpha$ (measured counter-clockwise from
the $x'$-axis in the plane of the sky)
\begin{equation}\label{eq:polarization_angle}
\alpha = \frac{1}{2}\textrm{atan2}\left(\frac{U}{Q}\right), 
\end{equation}
where the function ${\rm atan2}$ returns the appropriate quadrant of the
computed angle based on the signs of $Q$ and $U$, and the polarization fraction 
\begin{equation}\label{eq:fraction_sca}
p_{s,\infty}=\frac{I_p}{I_{s,\infty}},
\end{equation}
which can be compared directly with that given analytically in
eq.~\eqref{eq:fraction_limiting}.  

\begin{figure*}
  \includegraphics[width=0.49\textwidth]{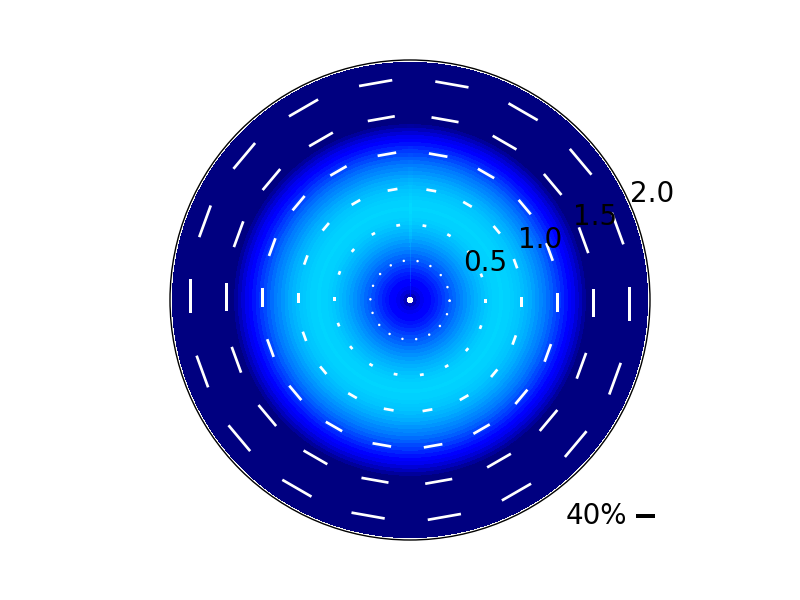}
  \includegraphics[width=0.49\textwidth]{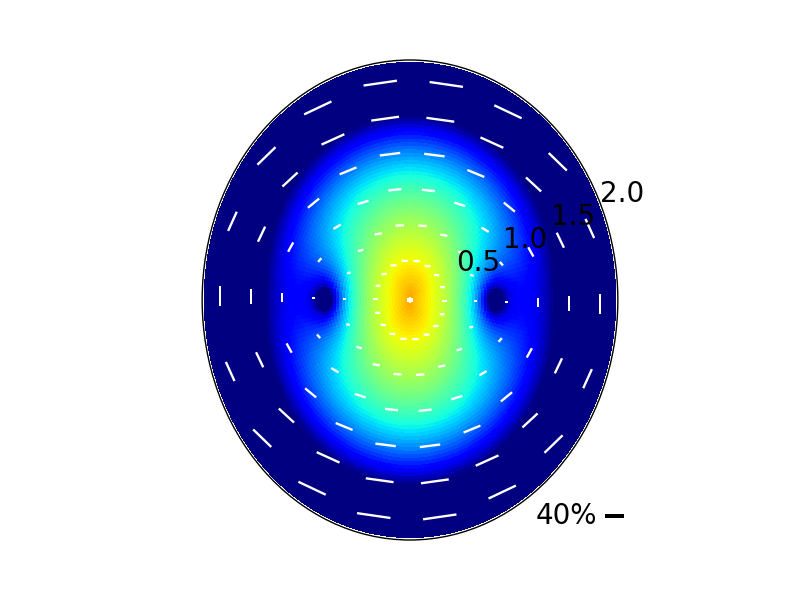}
  \includegraphics[width=0.49\textwidth]{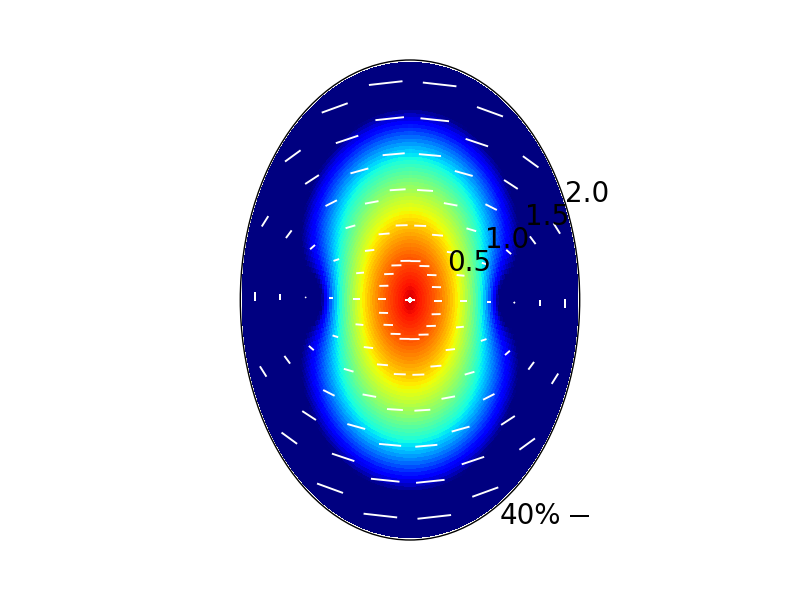}
  \includegraphics[width=0.49\textwidth]{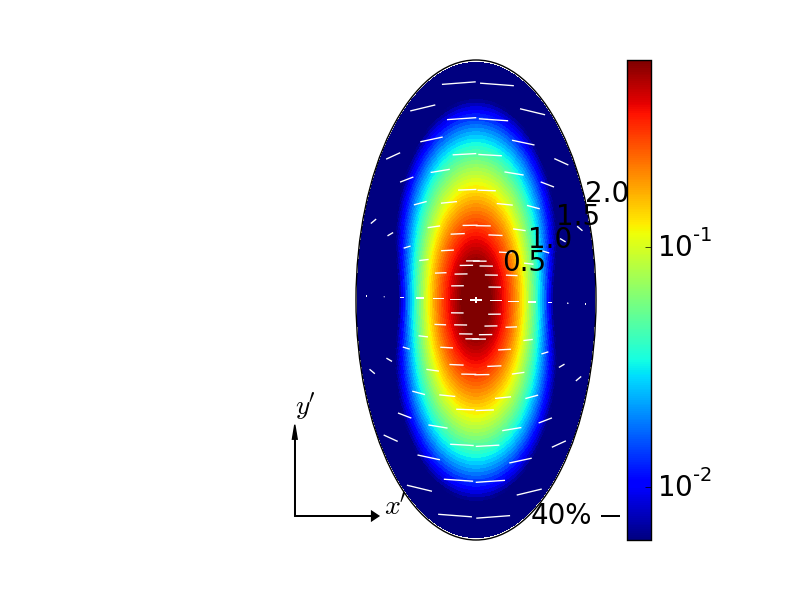}
  \caption{Effects of disk inclination on the polarization of 
    scattered millimeter radiation. Shown are the intensity of the 
    polarized radiation (color map,  in units of
    $2\Sigma_0^2\kappa_{abs}\kappa_{sca} \nu^2 k T_0/c^2$) and
    polarization vectors (line segments, with length proportional
    to the polarization fraction $p_{s,\infty}$ defined in
    eq.\eqref{eq:fraction_sca}), for $i=0^\circ$, $30^\circ$,
    $45^\circ$ and $60^\circ$. As the disk becomes more tilted, both
    the polarized intensity and polarization directions become more
    dominated by those induced by inclination. The elongation of the
    polarized intensity along the major axis and orientations of the
    polarization vectors along the minor axis in the high intensity region in
    the inclined cases are broadly consistent with the observed pattern in HL
    Tau (see Fig.~\ref{fig:obs}). The numbers 0.5, 1.0, 1.5 and 2.0 in each panel measure the
    de-projected distances from the center in units of the characteristic 
    radius $R_c$. The $x'$-axis and $y'$-axis defined in the plane of the
    sky are shown in the lower-right panel for reference.}
  \label{fig:illustration}
\end{figure*}

In Fig.\ref{fig:illustration}, we show the polarization vectors and 
the spatial distribution of the polarized intensity $I_p$ for 
$i=0^\circ$, $30^\circ$, $45^\circ$, and $60^\circ$. The intensity is
measured in units of $2\Sigma_0^2\kappa_{abs}\kappa_{sca} \nu^2 k 
T_0/c^2$, 
where $\Sigma_0$ and $T_0$ are the characteristic 
column density and temperature of the disk, and $\kappa_{abs}$ and 
$\kappa_{sca}$ the absorption and scattering opacity. In the face-on 
($i=0^\circ$) case, the fraction of polarization of the scattered light 
is zero at the center because the light to be scattered there comes 
isotropically along all azimuthal directions for the prescribed
axisymmetric disk. The radiation seen by the scatterer becomes 
more and more beamed in the radial direction as the radius 
increases, because of the drop in temperature and column density. 
As a result, the light is polarized in the azimuthal direction 
and the polarization degree increases outward. 
We will refer to the polarization induced by anisotropic radiation 
in the azimuthal direction in the face-on case as the ``intrinsic'' 
polarization. Note that although the polarization fraction $p_{s,\infty}$ 
can reach a value as high as $50\%$ or more near the outer edge, 
the total scattered polarized intensity $I_p$ is rather low. As a 
result, the intrinsic polarization from dust scattering can be easily
modified, indeed dominated, by the inclination-induced polarization.

When the disk is tilted away from the line of sight, both the polarized  
intensity and orientations of the polarization vectors change
drastically compared to the 
face-on case. As Fig.~\ref{fig:illustration} shows, the polarized 
intensity peaks at a ring in the face-on case, with the inward 
decrease caused by radiation becoming more isotropic in the 
azimuthal direction and the outward decrease from the exponential 
drop-off in column density. In contrast, the polarized intensity 
in the $i=30^\circ$ case peaks in the central region, as a result 
of the inclination-induced polarization. The polarization vectors 
in this (central) region (within $\sim 0.5 R_c$ of the origin) 
lie more or less along the minor axis, consistent with the 
analytic results for the inclination-induced polarization 
derived in the last subsection. Outside the central region, the 
polarization directions are broadly similar to those in the 
face-on case, indicating that the intrinsic polarization remains 
important there. A difference is that the axisymmetric polarization 
pattern in the face-on case becomes highly non-axisymmetric in 
this moderately tilted case, with both the polarized intensity and the 
polarization fraction significantly higher along the major 
axis than along the minor axis. 

As the disk tilt angle increases further, the polarization pattern becomes 
more dominated by that induced by inclination. Going from 
$i=30^\circ$ to $45^\circ$ to $60^\circ$, we see a clear trend for 
increasing polarized intensity in the central region, 
a larger fraction of the polarization vectors parallel to the 
minor axis, and more elongation of the polarized intensity 
along the major axis (see Fig.\ref{fig:illustration}). The 
elongation is a generic feature of the interplay between 
the intrinsic polarization and inclination-induced polarization. 
It provides a natural explanation for the distribution of 
polarized intensity observed in HL Tau\footnote{The observed direction
  of elongation does not lie exactly along the major axis, but is offset by
  a small angle of $\sim 10^\circ$ (see Stephens et al. 2014 and Fig.~\ref{fig:obs}). This offset
  is not explained in our model under thin-disk approximation, and may
  require full 3D models and/or additional physics, such as grain alignment.}, 
which has an inclination 
angle of $i\sim 45^\circ$ \citep{alma2015, kwon2011}.

\subsection{Interplay Between Intrinsic and Inclination-Induced
  Polarization}

\begin{figure}
  \includegraphics[width=0.49\textwidth]{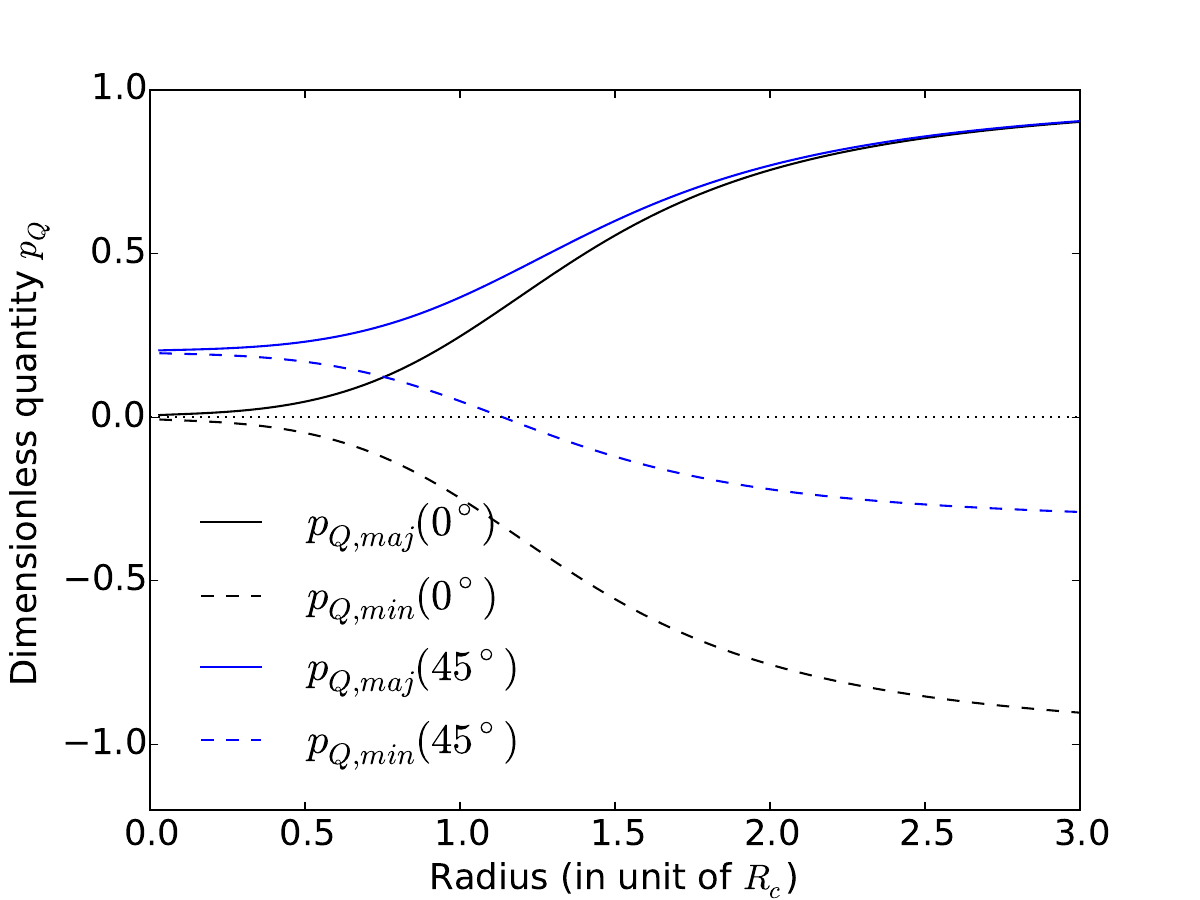}
  \caption{Dimensionless quantity $p_Q$ defined in
    eq.~\eqref{eq:pseudo-fraction}. Plotted are the distributions of
    $p_Q$ along the major and minor axes, denoted by the subscript
    ``maj'' and ``min'' respectively, as a function of de-projected
    distance from the origin for the face-on $i=0^\circ$ case (solid and
    dashed black) and the $i=45^\circ$ case (solid and
    dashed blue). 
    Its polarization fraction $|p_Q|$ approaches $20\%$ in the central
    region and, at larger distances from the origin, is substantially larger
    along the major axis than along the minor axis.}
  \label{fig:major_minor_inclination}
\end{figure}

To understand the interplay between the intrinsic polarization 
and inclination-induced polarization more quantitatively, we 
plot in Figure~\ref{fig:major_minor_inclination} the distribution 
of a dimensionless quantity 
\begin{equation}\label{eq:pseudo-fraction}
p_Q\equiv \frac{Q}{I_{s,\infty}}
\end{equation}
along the major and minor axes for the $i=45^\circ$ case. Since $U=0$ 
along the major and minor axes, this quantity is essentially 
the polarization fraction $p_{s,\infty}$ defined in
eq.\eqref{eq:fraction_sca}, except that it retains the sign of 
the Stokes parameter $Q$. A positive (negative) $p_Q$ means that 
the polarization is along the $x'$-axis ($y'$-axis) in the plane 
of the sky. 

In the face-on case ($i=0^\circ$), the intrinsic polarization 
fraction increases monotonically from zero to a large value
approaching unity as the distance from the origin increases 
along the major (or $y'$-) axis, with the polarization vectors 
parallel to the $x'$-axis in the plane of the sky. 
The quantity $p_Q$ is thus positive, 
as shown in Fig.~\ref{fig:major_minor_inclination} (solid 
black line). Along the minor axis, the fractional polarization 
is the same, but the polarization direction is along the 
$y'$-axis, with a negative $p_Q$ (dashed black line). For 
comparison, we plot in the same figure the numerically computed 
distribution of $p_Q$ along the major and minor 
axes as a function of the (de-projected) 
distance from the origin for the $i=45^\circ$ case. Clearly, 
the inclination has shifted the curves for the intrinsic ($i=0^\circ$) 
$p_Q$ along both major and minor axes upward, which is not 
surprising since inclination tends to produce polarization 
along the $x'$-direction (corresponding to a positive $Q$, 
see eq.\ref{eq:S_Qlimiting}). The amount of upward shift 
is different at different locations, however. In the central 
region where the intrinsic polarization fraction is relatively
low (within about 0.5~R$_c$), the shift is by $\sim 0.2$ 
along both the major and minor axes; it is the value 
expected from the analytic results for an azimuthally isotropic 
radiation field given in eq.~\eqref{eq:fraction_limiting}. 
The behavior in the central region where the intrinsic polarization is
weak is therefore easy to understand; it is dominated by 
the inclination-induced polarization. 
 
The behavior outside the central region where the intrinsic 
polarization fraction is higher is more complicated. It can 
be reproduced exactly, however, by the formula 
\begin{equation}\label{eq:fitting}
  p_Q = \frac{\sin^2 i}{2+\sin^2 i} + 
  \frac{p_{Q,in}}{(1+\sin^2i/2)[1+(1-p_{Q,in})\sin^2i/2]},
\end{equation}
along both the major and minor axes. Note that the first term 
on the right hand side is simply the inclination-induced 
polarization given by eq.~\eqref{eq:fraction_limiting}, and 
the second term is the intrinsic polarization fraction $p_{Q,in}$ 
(the face-on case) modified by the inclination. This formula can be
derived heuristically under the assumption that the radiation seen 
by the scatterer comes from only two directions: the $x$- and 
$y$-axes in the disk plane.  

In the limit where the intrinsic polarization fraction 
$|p_{Q,in}|\rightarrow 0$, we have 
\begin{equation}\label{eq:low_pol_limit}
p_Q \rightarrow \frac{\sin^2 i}{2+\sin^2 i}. 
\end{equation}
In the opposite limit where the intrinsic polarization fraction 
$|p_{Q,in}|\rightarrow 1$, we need to consider the major and 
minor axes separately. Along the major axis where $p_Q$ is 
positive, we have $p_{Q,in} \rightarrow 1$, which yields
\begin{equation}\label{eq:high_pol_limit_major}
p_Q \rightarrow 1. 
\end{equation}
Along the minor axis where $p_Q$ is negative, we have $p_{Q,in}
\rightarrow -1$, so that 
\begin{equation}\label{eq:high_pol_limit_minor}
p_Q \rightarrow - \frac{1-\sin^2 i}{1+\sin^2 i}, 
\end{equation}
which, for $i=45^\circ$ shown in
Fig.~\ref{fig:major_minor_inclination}, approaches $-1/3$. 
The difference in the asymptotic behavior of $p_Q$, 
particularly the polarization fraction $|p_Q|$,  
highlights one of the major differences between the 
polarization patterns along the major and minor axes. 
It becomes more extreme as the inclination angle $i$ 
approaches $90^\circ$. 

Another difference is that there exists a point of zero polarization 
on the minor axis across which $p_Q$ changes sign (or the 
polarization direction changes by $90^\circ$). This null point 
occurs at a location where 
\begin{equation}\label{eq:null}
p_{Q,in}=-\frac{\sin^2 i}{2-\sin^2 i},
\end{equation}
which has a value of $-1/3$ for $i=45^\circ$. Indeed, if the
inclination angle $i$ is known independently, one can in 
principle deduce the intrinsic value of $p_Q$ along the 
major and minor axes from the value of $p_Q$ through 
\begin{equation}\label{eq:inversion}
p_{Q,in}=\frac{p_Q - \sin^2 i/(2+\sin^2 i) }
{[(2-\sin^2 i) + p_Q \sin^2 i ]/(2+\sin^2 i) }. 
\end{equation}
However, it is difficult to infer the value of $p_Q$ 
from observation directly because it is the observed 
Stokes Q parameter normalized by the scattered intensity 
from the far-field, $I_{s,\infty}$, which cannot be measured 
directly. What can be measured is the total intensity, 
which we discuss next. 

\section{Total Intensity and Polarization Fraction}
\label{sec:application}

\begin{figure}
  \includegraphics[width=0.49\textwidth]{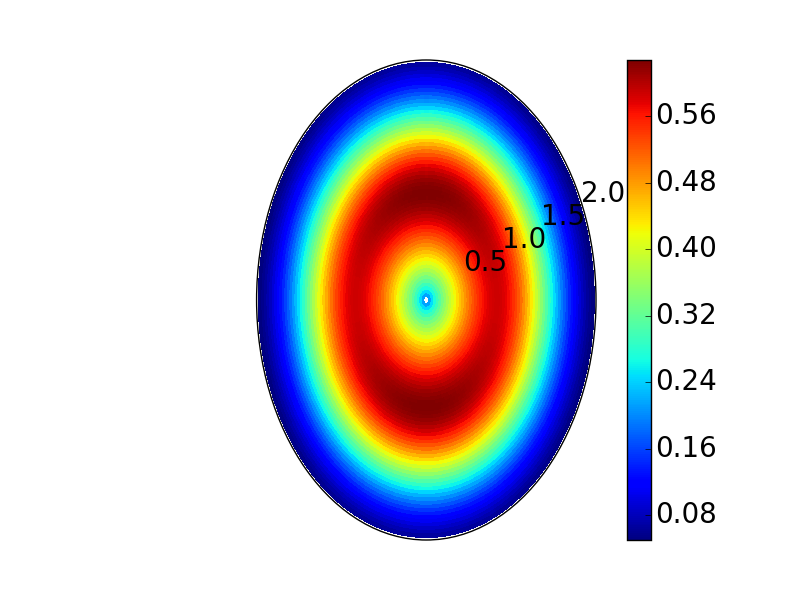}
  \caption{Ratio of the near-field and far-field contributions to the
    intensity of the scattered radiation, $I_{s,0}/I_{s,\infty}$, for the
  $i=45^\circ$ case. The far-field contribution $I_{s,\infty}$ dominates
  the near-field contribution $I_{s,0}$ everywhere, especially in the
  central region and near the outer edge. } 
\label{fig:Sratio}
\end{figure}

The total intensity $I$ of the radiation along the line of sight is the
sum of the direct dust emission $I_d$, the scattered radiation from 
near-field $I_{s,0}$, and the scattered radiation from far-field 
$I_{s,\infty}$. The source function for the near-field contribution at
a location ${\bf r}$ inside the disk can be estimated approximately
assuming isotropic incoming radiation from within a uniform sphere 
of radius $H$, the local scale height:
\begin{equation}\label{eq:S_0}
S_0\approx \frac{\nu^2 k \kappa_{abs} \Sigma({\bf r})T({\bf r})}{c^2},
\end{equation}
which is multiplied by the scattering optical depth $\Delta\tau_s \approx 
\kappa_{sca} \Sigma({\bf r})/\cos i$ to yield the intensity 
\begin{equation}\label{eq:I_0}
I_{s,0}\approx \frac{\nu^2 k \kappa_{abs}\kappa_{sca} \Sigma^2({\bf r})T({\bf
    r})}{c^2\cos i}.
\end{equation}
For the HL Tau disk model discussed in \S~\ref{disk_model}, the 
intensity of the scattered radiation $I_{s,0}$ from the near-field is 
weaker than that from the far-field $I_{s,\infty}$ everywhere for 
$i=45^\circ$, as can be seen from Fig.~\ref{fig:Sratio}, where 
the ratio of the two is plotted. The ratio peaks in a ring between 
$\sim 0.5 R_c$ and $\sim 1 R_c$ (where $R_c$ is the characteristic 
radius of the disk), with a maximum value of $\sim 60\%$. Near 
the peak, the unpolarized scattering 
intensity $I_{s,0}$ reduces the polarization fraction by a 
factor of $\sim 1.6$, from $\sim 20\%$ to $\sim 12\%$. Going 
outward from the ring, the ratio drops rapidly because the 
near-field intensity $I_{s,0}$ is determined by the local column 
density, which drops off exponentially with radius, whereas 
the far-field intensity $I_{s,\infty}$ is determined globally, 
including contributions from the bright central region that
decrease with radius more slowly than exponential. Inside 
the ring, the ratio decreases with decreasing radius 
because of a smaller scale height $H$, which decreases the 
size of the region where the incoming radiation for the 
near-field source function $S_0$ comes from relative to that 
for the far-field source function $S_\infty$. In any case, the 
polarization fraction of the scattered radiation remains 
high, of order $10\%$ or more, after both the near- and 
far-field contributions are taken into account. It is much 
higher than observed in HL Tau (of order $1\%$), and needs 
to be further reduced, by the unpolarized direct dust 
emission\footnote{Direct emission can be polarized, due to, for
  example, grain alignment. Polarized direct emission needs to be 
included in a more complete model (as discussed in
\S~\ref{sec:discussion}).}. 

The source function for the direct emission is simply the Planck
function $B_\nu(T)\approx 2\nu^2 kT/c^2$. The optical depth for
dust absorption along the line of sight passing through a 
location ${\bf r}$ in the disk is 
$\Delta \tau_a\approx \kappa_{abs} \Sigma({\bf r})/\cos i$ under 
the thin disk approximation. Together, they yield the intensity
for the direct emission 
\begin{equation}\label{eq:direc_emission}
I_d\approx \frac{2\nu^2 k \kappa_{abs} \Sigma({\bf r}) T({\bf r})}{
  c^2 \cos i}. 
\end{equation} 
This estimate allows us to evaluate the ratio of the intensities of the 
scattered and direct emission: 
\begin{equation}\label{eq:I_ratio}
  \begin{split}
\frac{I_s}{I_d}&=\frac{I_{s,0}+I_{s,\infty}}{I_d}
\approx \kappa_{sca}\Sigma_0\frac{\Sigma({\bf
    r})}{\Sigma_0} \times\\
    &\left[\frac{1}{2} + \frac{3}{16\pi} \int_0^{2\pi} d\phi
    (1+\sin^2i\cos^2\phi)\int_H^\infty \frac{dl}{l}\frac{\Sigma({\bf
        r_l})}{\Sigma({\bf r})} \frac{T({\bf r_l})}{T({\bf r})}
  \right].  
  \end{split}
\end{equation}
Note that the ratio $\Sigma({\bf r})/\Sigma_0$ and the second term inside the square bracket are dimensionless
quantities that depend only on the shape of the column density and
temperature profiles. The overall scaling is set by the characteristic
scattering optical depth $\Delta\tau_{s,c}=\kappa_{sca}\Sigma_0$. 
In order to reduce the high polarization
fraction of the scattered radiation at 1.3~mm to the observed value 
of about $1\%$, we need a rather small value of $\Delta\tau_{s,c} 
\approx 0.07$, so that the scattered radiation is heavily diluted by 
the unpolarized direct emission. 

\begin{figure}
  \includegraphics[width=0.49\textwidth]{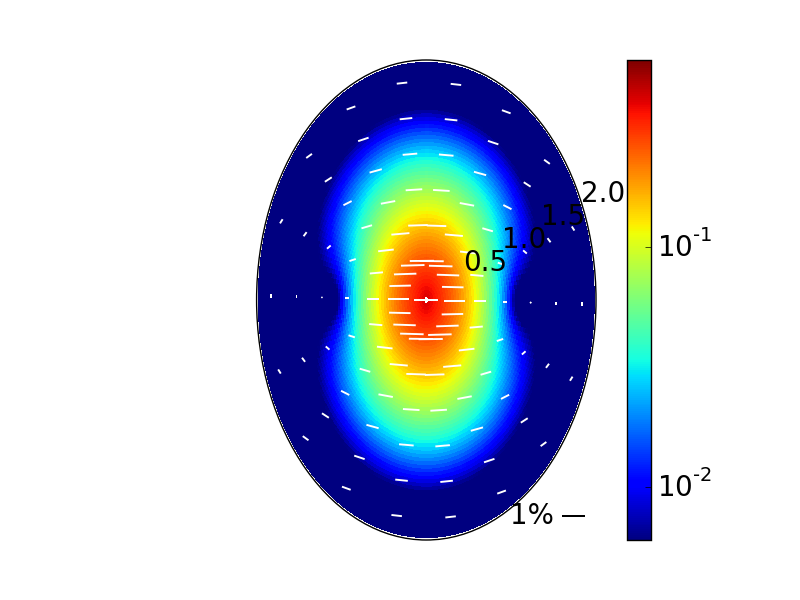}
  \caption{Model for the polarization of HL Tau disk. Plotted are the
    polarized intensity (color map, in units of $2\Sigma_0^2\kappa_{abs}\kappa_{sca} \nu^2 k 
T_0/c^2$) and
    polarization vectors (line segments, with length proportional
    to the polarization fraction of the total intensity) for
    $i=45^\circ$. A characteristic scattering optical depth
    $\Delta\tau_{s,c}\approx 0.07$ is needed to bring the high polarization
  fraction of tens of percent relative to the scattered intensity
  (shown in the lower-left panel of Fig.~\ref{fig:illustration}) down
  to the observed level of $\sim 1\%$ relative to the total intensity.}
  \label{fig:final_map}
\end{figure}

We stress that the inclusion of the unpolarized radiation $I_{s,0}$
and $I_d$ changes neither the polarized intensity nor the polarization
direction shown in Fig.~\ref{fig:illustration} (the lower-left panel). 
What is changed is the polarization fraction. In 
Fig.~\ref{fig:final_map}, we plot the distribution of the polarized
intensity for the $i=45^\circ$ case, as in the lower-left panel of 
Fig.~\ref{fig:illustration}, but with the length of the overlaid 
polarization vectors scaled by the new polarization fraction (relative
to the total intensity). This figure represents our final model for the
HL Tau disk. It has three features that are broadly consistent with 
observations: (1) the region of high polarized intensity is elongated 
along the major axis, (2) the polarization vectors in this region are 
nearly parallel to the minor axis, and (3) the polarization fraction 
in the region is about $1\%$. Along the ridge of detectable polarized
intensity, the observed polarization fraction appears to be somewhat 
higher toward the edge of the disk (at a distance of $\sim R_C\approx 
80$~AU) than near the center, although it is unclear how significant 
the trend is in view of the low polarized intensity near the edge. 
This trend was present in our model when 
only the scattered radiation was considered (see
Fig.~\ref{fig:illustration} and \ref{fig:major_minor_inclination}), 
but was washed out by the total intensity in Fig.~\ref{fig:final_map}.

The semi-analytic theory that we have developed so far under the assumption of 
geometrically thin disk, optically thin emission, and only Rayleigh 
scattering is independent of the detailed properties of dust 
grains. This independence makes the broad agreement between our 
model and the main polarization features observed in HL Tau 
rather robust.  In the next section, we will try to put constraints 
on the grain size distribution in the HL Tau disk, which is much 
more uncertain.  

\section{Implications on Grain Size}
\label{sec:implications}

\subsection{Scattering Opacity and the Need for Large Grains} 
\label{subsec:grain}

The main free parameter of the final model for the HL Tau disk
polarization discussed 
in \S~\ref{sec:application} is the characteristic scattering optical
depth $\Delta\tau_{s,c}$, which controls the polarization fraction. 
In order to produce the observed polarization fraction of $\sim 1\%$,
a value of $\Delta\tau_{s,c}\approx 0.07$ is required. This value  
yields a scattering opacity $\kappa_{sca}\approx 
10^{-3}$~cm$^2$~g$^{-1}$ (cross section per unit total, rather than
dust, mass) at 1.3~mm (the wavelength of the HL Tau disk  
polarization observation), using the characteristic column density 
$\Sigma_0=68$~g~cm$^{-2}$ from the best-fit disk mass of
$0.13$~M$_\odot$ of \cite{kwon2011}. This scattering opacity can 
put constraints on the grain size distribution, although they 
depend on the dust composition, which is uncertain. As an
illustration, we consider the model of dust grains adopted by 
\cite{kataoka2015a}, which are spheres with a mixture of silicate (8\%), 
water ice (62\%) and organics (30\%). All fractional abundances are in volume
and are taken from \cite{pollack1994}.
We assume a canonical gas-to-dust mass ratio of 100,
and use the Mie theory to calculate the absorption and scattering 
opacities \citep{bh1983}. The inferred scattering opacity corresponds to a grain 
radius $a=37\rm\, \mu m$ for grains of a single size. For the MRN-type  
power-law distribution $n(a) \propto a^{-3.5}$ \citep{mrn1977}, we obtain 
a maximum grain 
size of $a_{max}=72\rm\, \mu m$. The increase of this maximum over 
the single size case comes from averaging over the grain size. 
In both cases, the dimensionless parameter $x=2\pi a/\lambda 
\ll 1$, so that the Rayleigh limit used for treating the 
scattering in the previous sections is self-consistent (see 
Fig.~\ref{fig:preversal} below).  
The maximum size inferred for the grains responsible for the scattered
dust emission in the HL Tau disk is much larger than that of the
grains in the diffuse interstellar medium. This is consistent with 
other lines of evidence for grain growth in disk environments (e.g.,
\citealt{perez2012, alecian2013, testi2014}).

  We note that \cite{kataoka2015b} independently modeled the HL Tau disk
polarization using dust scattering through Monte Carlo radiative
transfer simulations. They obtained disk polarization patterns that are
very similar to ours. They inferred a maximum grain size that ranges 
from 70 $\mu$m to 350 $\mu$m, which is broadly consistent with our value 
of 72 $\mu$m.

In summary, to reproduce the $\sim 1\%$ polarization fraction 
observed in the disk of HL Tau through dust scattering, 
the grains must have grown to tens of microns (the exact 
value depends on the assumed grain size 
distribution and composition). However, this picture is 
complicated by the opacity spectral index $\beta$ inferred 
for HL Tau, as we discuss next. 

\subsection{Opacity Spectral Index $\beta$ and the Need for Larger Grains?}
\label{subsec:larger_grains}

\cite{kwon2011} obtained a best-fit value $0.73$ for the spectral index
$\beta$ of the dust opacity $\kappa_{abs}\propto \nu^\beta$ for the HL
Tau disk based mostly on CARMA observations at 1.3 
and 2.7 mm. It is in agreement with the spatially averaged value 
obtained from ALMA observations from 0.87 to 2.9 mm \citep{alma2015}. 
This value is significantly lower than the 
typical ISM value of $\beta\sim 1.5-2$. The difference is 
usually taken as evidence for grain growth to millimeter 
size or larger \citep{testi2014}, 
although other interpretations are possible. 
For example, \cite{ricci2012} showed that a value of 
$\beta\sim 1$ or lower can be obtained without mm/cm sized 
grains if part of the disk is optically thick. Some support 
for this possibility is provided by the spatially resolved 
distribution of $\beta$ derived from the ALMA data, which 
shows $\beta\sim 0$ indicative of optically thick emission 
at the central continuum peak and two rings (B1 and B6, 
\citealt{alma2015},
see their Fig.~3). Another possibility is that the index 
$\beta$ is sensitive to not only the size but also the shape 
of the grains. Indeed, \cite{verhoeff2011} was able to 
reproduce the spectral energy distribution (SED) of the disk 
of HD 142527 (with $\beta\sim 1$ in the millimeter regime)  
with irregular grains of sizes up to only 2.5~$\rm \mu m$; the 
grain shape was treated with the distribution of hollow 
spheres \cite{min2005}. The grains inferred in our model 
of dust scattering-induced polarization for the HL Tau disk 
have a significantly larger maximum size (of order tens of 
microns). They may still be able to reproduce the observed 
(averaged) opacity spectral index of $\beta\sim 0.73$ if the 
grains are irregular and/or part of the disk is optically 
thick. Detailed exploration of this possibility is beyond the 
scope of the present work. 

If large, mm/cm sized, grains are responsible for the relatively 
low value of $\beta$ observed in the HL Tau disk, it is natural 
to ask whether they can produce a polarization pattern that 
matches the observed one through scattering. It is 
unlikely, because the key to producing the observed pattern is 
the polarization degree of the scattered light peaking near 
$90^\circ$ (as in the Rayleigh limit), and this requirement is not 
satisfied for mm/cm sized grains. For example, for the grain model 
adopted by \cite{kataoka2015a}, the polarization degree (defined 
as the ratio of the two elements in the scattering matrix, 
$-Z_{12}/Z_{11}$, which is essentially the polarization fraction but can be
either positive or negative) is nearly zero at 0.87 mm for all scattering 
angles except around $135^\circ$, where it reaches a (negative) ``peak''
value of $\sim -0.2$ for $a_{max}=1$~mm and 1~cm (see the right 
panel of their Fig.~2). 
The negative value is known as the polarization reversal 
(e.g., \citealt{murakawa2010, kw2014}) which, 
together with the shift of the polarization ``peak'' away from 
$90^\circ$, is expected to produce a polarization pattern very 
different from the Rayleigh scattering case. 

\begin{figure}
  \includegraphics[width=0.49\textwidth]{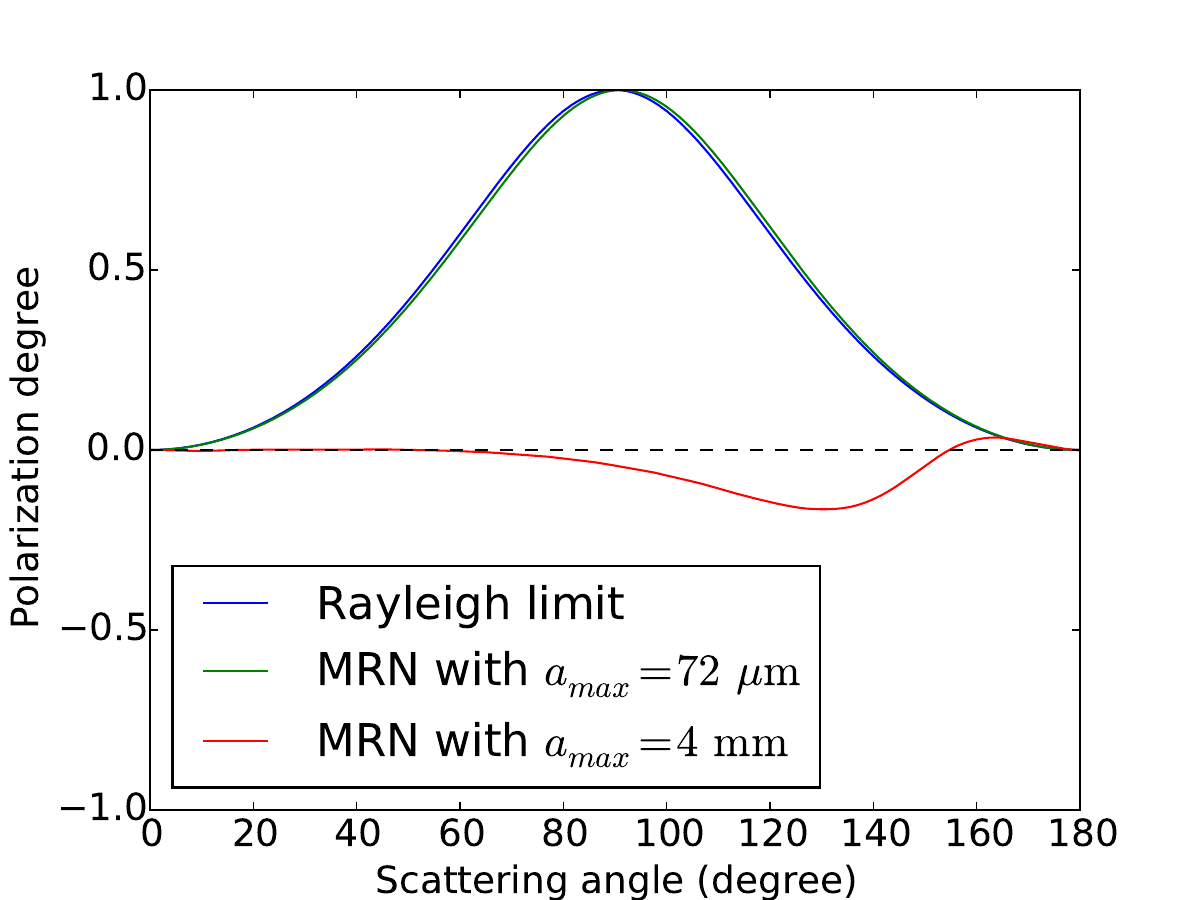}
  \caption{Comparison of the distributions of the degree of
    polarization (see text for definition) as a function of scattering angle for three cases:
    the Rayleigh limit (blue line) and two cases with MRN-like
    power-law grain size distribution with $a_{max}=72\; {\rm \mu m}$
    (green) and 4~mm (red). The Rayleigh and $a_{max}=72\; {\rm \mu
      m}$ cases are almost indistinguishable. The negative ``peak''
    around $130^\circ$ in the $a_{max}=4$~mm case is an example of
    the so-called ``polarization reversal,'' which may provide a way
    to probe large, mm/cm sized, grains through scattering-induced
    polarization.}
  \label{fig:preversal}
\end{figure}

As an illustration, we repeat the computation of the 
scattering-induced polarization at $\lambda=1.3$~mm 
in \S~\ref{sec:application}, but with an MRN-type power-law 
size distribution up to $a_{max}=4$~mm (instead of 72~$\rm \mu m$), 
using the dust model of \cite{kataoka2015a} and Mie theory. 
The maximum grain size is chosen such that $a_{max}\approx 3 
\lambda$, which is roughly the minimum value required to yield an 
opacity spectral index of $\beta \sim 1$ according to \cite{draine2006}. 
The distribution of the polarization degree with scattering
angle in this case is shown in Fig.~\ref{fig:preversal}. It is very 
similar to that obtained by \citeauthor{kataoka2015a} at 0.87 mm, except 
that the ``peak'' is slightly lower (-0.17) and is shifted to a 
slightly smaller angle of $\sim 130^\circ$.   

\begin{figure}
  \includegraphics[width=0.49\textwidth]{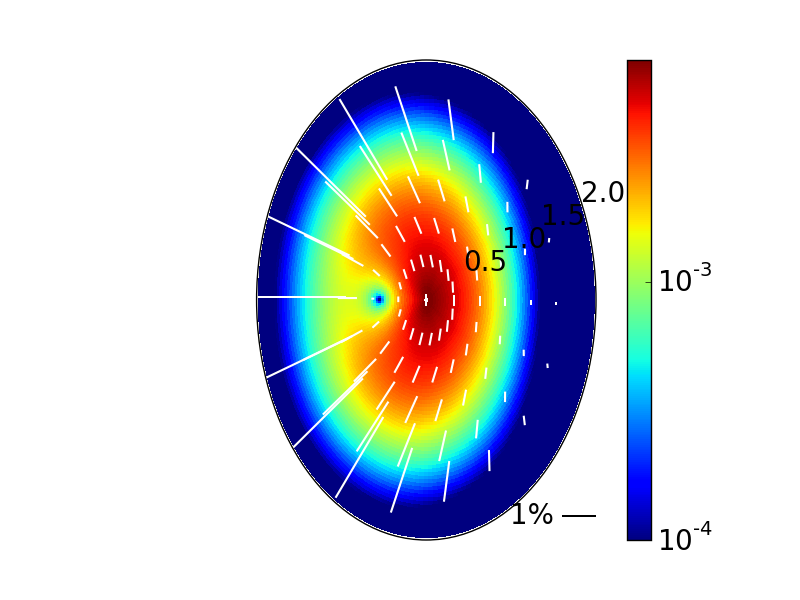}
  \caption{Scattering-induced polarization by large grains. As in
    Fig.~\ref{fig:final_map}, plotted are the
    polarized intensity (color map) and
    polarization vectors (line segments, with length proportional
    to the polarization fraction). Note the strong asymmetric with
    respect to the major axis in both the polarized intensity and the
    polarization vectors. The polarization along the major axis in the
    central region is due to polarization reversal, which may be a 
    robust indicator of scattering by large, mm/cm-sized, grains. The
    near side of the disk is on the right. }
  \label{fig:large_grain}
\end{figure}

In Fig.~\ref{fig:large_grain}, we plot the distribution of the 
polarized intensity together with polarization vectors for the 
large grain case of $a_{max}=4$~mm. There are several features 
that are worth 
noting. First, unlike the Rayleigh scattering case, 
the polarized intensity is no longer symmetric 
with respect to the major axis. This is because large, mm/cm sized, 
grains preferentially scatter light in the forward direction (e.g.,
\citealt{bh1983}), 
making the side of the disk closer to the observer (the right half) 
brighter. The polarization fraction is, however, higher on 
the far side (especially toward the outer part of the disk)  
because the polarization degree of the scattered light is 
higher for backward scattering than for forward scattering 
(see Fig.~\ref{fig:preversal}).  
The most striking difference between this case and the 
Rayleigh scattering case shown in Fig.~\ref{fig:final_map} lies 
in the polarization direction. The difference comes from 
the polarization reversal in the large grain case, which yields 
an intrinsic (or face-on) polarization direction in the radial 
(as opposed to azimuthal) direction and an inclination-induced 
polarization along the major (rather than minor) axis. The 
interplay between the intrinsic and inclination-induced 
polarization leads to polarization directions in the region 
of high polarized intensity (the most easily observable 
part) completely different from those observed in HL Tau (see 
Fig.~\ref{fig:obs}).  

We are therefore left with an interesting conundrum. The 
polarization pattern in the HL Tau disk is suggestive 
of Rayleigh scattering by relatively small dust grains 
(although still much larger than the typical ISM grains), 
but such grains may have difficulty reproducing 
the observed opacity spectral index $\beta$ ($\lesssim 1$). 
The index can be reproduced more easily with larger, mm/cm 
sized, grains, but it is difficult to generate the observed 
polarization pattern with such grains through scattering. 
It is conceivable that there are two populations of 
dust grains, with one responsible for polarization, the other 
for $\beta$. The two populations do not have to be located 
co-spatially in the disk; for example, large grains responsible 
for the bulk of the unpolarized continuum (and thus $\beta$) 
may have settled close to the midplane, whereas smaller grains 
that dominate the polarized millimeter radiation may remain 
floating higher up above the midplane 
(e.g., \citealt{dd2004, tanaka2005, balsara2009} ). If 
this speculation turns out correct, polarized emission in 
millimeter would provide a powerful probe of not only 
grain growth, but also the expected vertical stratification 
of grain sizes, especially in conjunction with observations 
of optical/IR polarization, which probe even smaller, 
micron-sized, grains that are higher up still above the 
disk midplane.
 
\section{Other Sources and Future Directions}
\label{sec:discussion}

As mentioned in \S~\ref{sec:intro}, spatially resolved polarized
sub-mm/mm emission was observed in two other sources besides 
HL Tau: IRAS 16293-2422B \citep{rao2014} and L1527 \citep{segura-cox2015}, 
using SMA and CARMA respectively. We discuss whether the 
polarization in these two sources is compatible with an origin in 
dust scattering. 

The L1527 disk is nearly edge-on, with an inclination angle $i\approx
90^\circ$. Its 1.3 mm dust emission was observed to be polarized at $\sim 2.5\%$ 
level, with a direction roughly perpendicular to the disk (i.e., along
the minor axis). This 
pattern is consistent with that expected from dust grains aligned 
by a predominantly toroidal magnetic field \citep{segura-cox2015}. 
It is also the pattern 
expected of scattering by relatively small dust grains in the 
Rayleigh limit because the polarization induced by disk inclination
is along the minor axis (as illustrated vividly in the lower right 
panel of Fig.~\ref{fig:illustration}). Indeed, the
fraction of the scattered
radiation that is polarized due to disk inclination increases 
with the inclination angle, reaching a maximum value of $1/3$ for
edge-on disks, as we showed analytically in \S~\ref{sec:isotropic} 
(see eq.~\eqref{eq:fraction_limiting}). This trend makes it more 
likely for L1527-like disks to show scattering-induced polarization 
than face-on disks. Indeed, the median polarization fraction of L1527
is the highest among the three sources with spatially resolved sub-mm/mm 
polarized emission so far. However, whether such polarization can 
actually be observed 
or not depends on the polarized intensity, which in turn depends on the 
total intensity and the polarization fraction (relative to the 
total intensity). The latter is sensitive to the scattering opacity
$\kappa_{sca}$, which depends on the dust properties, especially 
the grain size, which can vary greatly from one source to another. 
To produce the observed polarization fraction of 
$\sim 2.5\%$ in L1527 through scattering, the grains must have 
grown well beyond $\rm \mu m$ size; otherwise, the scattering opacity 
would be too small. The scattering cannot be dominated by large, 
mm/cm sized grains either. Such grains would produce polarization 
along, rather than perpendicular to, the edge-on disk because of 
polarization reversal (see \S~\ref{subsec:larger_grains} and 
Figs.~\ref{fig:preversal} and \ref{fig:large_grain}). Large cm-sized grains 
are inferred in the L1527 disk from the small opacity spectral 
index $\beta\sim 0$ \citep{tobin2013}. If the observed polarization 
is due to scattering by sub-mm sized grains, then two grain populations 
may again be needed, as in the HL Tau case discussed in
\S~\ref{subsec:larger_grains}.

The disk in IRAS 16293-2422B appears to be nearly face-on 
\citep{rodriguez2005, zapata2013}. Its 0.88 mm 
emission was observed to be polarized at $\sim 1.5\%$ level, 
with the polarization directions showing a swirling 
pattern that is neither strictly radial nor purely azimuthal 
\citep{rao2014}. \citeauthor{rao2014} showed that the pattern is 
broadly consistent with the polarized emission expected from 
grains aligned by a magnetic field that is warped into a spiral 
configuration by disk rotation, although the rotation is hard to
measure directly because of the face-on orientation. 
It is inconsistent with the 
simplest dust scattering model for axisymmetric face-on 
disks, where the polarization directions are expected to be
perfectly azimuthal, as illustrated in the upper left panel 
of Fig.~\ref{fig:illustration}. However, the observed 
polarized intensity is arc-shaped which, for face-on systems, 
requires intrinsically non-axisymmetric disk models, such 
as those constructed by \cite{kataoka2015a}. Indeed, the 
observed intensity distribution is reminiscent of the 
lopsided ring models of \citeauthor{kataoka2015a} (see their Fig.~6 
and 7), although the polarization direction in the models 
changes sharply from radial in the inner part of the ring 
to azimuthal in the outer part. It is conceivable that 
dust scattering models with more complicated disk structures, 
such as spiral arms, may match the observation better, 
but this remains to be demonstrated. One worry is that the 
dust emission in this source may be optically thick at 
sub-mm wavelengths \citep{zapata2013}, which would reduce 
the degree of anisotropy in the unpolarized radiation to 
be scattered, and thus the degree of polarization in the 
scattered radiation \citep{kataoka2015a}. Another is that 
the polarized emission detected in this deeply embedded 
source may be contaminated by the protostellar envelope. 
Higher resolution ALMA polarization observations should 
become available in the near future. They will provide more 
stringent tests of the dust scattering model of polarized 
mm/sub-mm emission.  
 
As stressed by \cite{kataoka2015a}, polarized radiation at mm/sub-mm
wavelengths provides a powerful probe of grain growth, if it 
is produced by dust scattering. A robust prediction of the 
scattering model is that large, mm/cm-sized, grains should produce
millimeter polarization along the major axis of an inclined disk due
to polarization reversal, especially in the central region where 
the intrinsic (face-on) polarization is expected to be weak and 
the observed polarization pattern is more easily dominated by that 
induced by disk inclination (see Fig.~\ref{fig:large_grain} for 
an illustration). This effect should be searched for with high 
resolution ALMA observations, especially in high-inclination
systems. Another way to probe large grains is to observe 
polarization at longer, centimeter, wavelengths using, for 
example, JVLA and the future SKA. At such wavelengths, the 
scattering by mm-sized grains would still be in the Rayleigh 
regime, with a high degree of polarization of the scattered 
light. Given the sensitive dependence of the scattering opacity 
$\kappa_{sca}$ (which controls the polarization fraction of the 
total intensity) on the grain size relative to the wavelength, 
it is important to carry out high-resolution polarization 
observations at multiple wavelengths to determine whether 
the polarization is indeed due to dust scattering and, if yes, 
to constrain the grain size distribution. With the polarization
capability of JVLA, and that of ALMA coming online soon, the 
prospect for using sub-mm/mm/cm polarization to probe grain 
properties in disks is bright on the observational side.     

On the theoretical side, much work remains to be done. In this 
paper, we have limited our treatment to the simplest case of 
optically and geometrically thin disk, to bring out the 
essential features of the disk inclination-induced polarization 
transparently through a semi-analytic model. As noted earlier, there 
is indication from the spatial distribution of the opacity spectral 
index $\beta$ that part of the HL Tau disk is optically thick at 
sub-mm/mm wavelength, including the central continuum peak, B1 and 
B6 rings, at radii of $\sim 0$, 20, and 81~AU, respectively \citep{alma2015}.
Since the B6 ring is rather narrow, 
and the bulk of polarized emission is detected interior to it, it 
should not affect our HL Tau model much. The continuum peak and B1 
ring are not resolved by the polarization observation, but they can 
potentially lower the polarization fraction in the central pixel 
relative to the outer part, which is expected to bring the dust 
scattering model presented in \S~\ref{sec:application} and 
Fig.~\ref{fig:final_map} into closer agreement with
observation\footnote{The polarization fraction of millimeter emission 
in the central region can also be lowered if the grains there grow to 
centimeter sizes, which can contribute significantly to the total 
but little to the polarized intensity.}. 
The effects of optically thick regions, as well as substructures 
on the disk such as rings and gaps, need to be quantified in 
future calculations in order to compare with higher resolution 
ALMA polarization observations that should become available soon. 

Another future improvement is to relax the thin disk approximation. 
While the approximation is adequate for large grains that have 
settled close to the disk midplane, it would be less so for smaller 
grains that remain higher up above the midplane. It is likely that 
there is a gradient in dust grain concentration and size distribution 
in all three (radial, vertical and azimuthal) directions through 
grain growth, inward drift, vertical settling, and trapping (see 
\citealt{testi2014} for a recent review, and \citealt{perez2012} for an
example of gradient in grain size). Such gradients should be 
taken into account in more complete models of scattering-induced 
polarization, perhaps using 3D radiative transfer codes such as 
RADMC3D. As noted earlier, in HL Tau and L1527, the observed opacity 
spectral index $\beta$ and polarization pattern in millimeter, if 
originating from dust scattering, appear to require two grain  
populations of different size distributions. Whether they can arise 
naturally through grain evolution in the disk warrants further 
investigation. In addition, a complete model of disk polarization 
at sub-mm/mm/cm will need to include both the polarized emission 
by aligned, non-spherical grains 
(particularly grains of tens of microns in size or larger, although 
it is unclear whether such grains can be aligned with respect to the 
magnetic field or not, as discussed in \S~\ref{sec:intro})
and dust scattering, and possible 
interplay between the two.  

\section{Conclusions}
\label{sec:conclusion}

Motivated by the recent spatially resolved millimeter polarization
observations of the HL Tau disk, we have developed a simple
semi-analytic model for the dust scattering-induced polarization 
in the limit of optically and geometrically thin  
disk and Rayleigh scattering, with an emphasis on the effects of 
the disk inclination to the line of sight. The main results are 
summarized as follows: 

1. We developed an efficient approximate method for computing disk polarization
from dust scattering by dividing the source region of the millimeter 
radiation to be scattered at a location inside the disk into 
two conceptually distinct parts: a near-field region centered 
on the location with a size comparable to the local dust scale-height, 
and a far-field region outside. Radiation from the near-field region
is more or less isotropic, and does not contribute significantly to
the polarization of the scattered light. Radiation from the far-field
region is concentrated toward the disk plane. It is strongly 
polarized after scattering in an inclined disk. The polarization 
fraction of the scattered light increases with the inclination 
angle, reaching a maximum value of $1/3$ for edge-on disks if 
the incoming radiation to be scattered is azimuthally isotropic 
in the disk plane (eq.~\eqref{eq:fraction_limiting}). The polarization
induced by disk inclination is parallel to the minor axis. It can
easily dominate the intrinsic polarization of the disk in the 
face-on view (see Fig.~\ref{fig:illustration}).

2. We developed a simple model for the polarization of the HL Tau
disk, based on the\citep{kwon2011} model of disk physical 
structure and polarization induced by a disk inclination of $45^\circ$ 
(see Fig.~\ref{fig:final_map}). The 
model naturally reproduces two main features of HL Tau: (1) the region
of high polarized intensity is elongated along the major axis, and (2)
the polarization vectors in this region are roughly parallel to the
minor axis.  Both are the consequences of a simple geometric 
effect: only the radiation propagating along the major axis of a
tilted disk would be scattered by 90$^\circ$ to reach the observer and 
be maximally polarized, with a 
polarization direction along the minor axis in the plane of the 
sky. The broad agreement is robust because it does not depend on 
the detailed properties of dust grains (which are uncertain) as long 
as the scattering is in the Rayleigh limit.   
It provides support for the millimeter polarization 
observed in this particular case originating at least in part 
from dust scattering, although polarized emission from magnetically 
aligned dust grains cannot be ruled out, especially if the disk 
field is more complex than toroidal. 

3. For the other two cases with observed mm/sub-mm polarization, L1527
and IRAS 16293-2242B, the situation is less clear. The observed
polarization pattern in the nearly edge-on disk of L1527 is compatible with
that expected of either toroidal field-aligned grains or dust
scattering. The pattern observed in the possibly face-on disk of 
IRAS 16293-2242B is more consistent with that expected of grains
aligned by a rotationally warped magnetic field than with the simplest
case of dust scattering. Higher resolution observations of more
disks with different inclination angles are needed to better 
differentiate the grain-alignment and dust-scattering models. The 
observational situation should improve drastically with ALMA and 
JVLA.   

4. To reproduce the polarization fraction of $\sim 1\%$
observed at 1.3 mm in the HL Tau disk, a maximum size of tens 
of microns is needed for the scattering grains. Such grains 
are generally thought to be too
small to produce an opacity spectral index $\beta$ of order 1 or less 
that is observed in HL Tau and other sources; larger, mm/cm sized 
grains may be needed. However, mm/cm sized grains tend to produce
polarization parallel (rather than orthogonal) to the major axis 
due to polarization reversal (see Figs.~\ref{fig:preversal} and
\ref{fig:large_grain}), which is not observed in HL Tau; nevertheless,
this pattern should be searched for in other sources as a robust 
indicator for large grains. In any case, the dust scattering model 
for polarization and the relatively low $\beta$ produce an interesting 
conundrum that needs to be resolved in the future, perhaps with 
more complete models that include grain evolution and 3D radiative 
transfer, as well as polarized direct emission from aligned 
grains. Such models, together with the high resolution ALMA/JVLA 
polarization observations that will soon become available, should 
make it possible to disentangle the contributions of grain alignment
and dust scattering to the disk polarization, which is needed in order
to provide robust constraints on the magnetic field that is generally
thought to be crucial to the dynamics and evolution of protoplanetary 
disks and/or the grain growth that may eventually lead to planet 
formation. 

\section*{Acknowledgements}

We thank Phil Arras, Dom Pesce, Scott Suriano, and Laura P\'{e}rez  
for helpful discussion. LWL acknowledges support by NSF
AST-1139950. HFY and ZYL are supported in part by NSF
AST-1313083 and NASA NNX14AB38G.

%%%%%%%%%%%%%%%%%%%%%%%%%%%%%%%%%%%%%%%%%%%%%%%%%%

%%%%%%%%%%%%%%%%% APPENDICES %%%%%%%%%%%%%%%%%%%%%

%%%%%%%%%%%%%%%%%%%%%%%%%%%%%%%%%%%%%%%%%%%%%%%%%%

% Don't change these lines
\bsp	% typesetting comment
\label{lastpage}
\end{document}